\documentclass[11pt,twoside]{article}
\usepackage{newpasp,epsf}
\def\edcomment#1{\iffalse\marginpar{\raggedright\sl#1\/}\else\relax\fi}
\reversemarginpar

\begin{document}
\title{Anisotropic inverse Compton scattering in radio galaxies and 
particle energy distribution}
 \author{Gianfranco Brunetti and Giancarlo Setti}
\affil{Dip. Astronomia, Univ. di Bologna, 
via Ranzani 1, 40127 Bologna, Italy}
%\author{Ima Co-Author}
\affil{Ist. di Radioastronomia del CNR, via Gobetti 101, 
40129 Bologna, Italy}

\begin{abstract}
In this talk we briefly review the inverse Compton (IC) scattering
of nuclear photons in the lobes of radio galaxies and quasars.
We concentrate on the possibility to test this model 
with the {\it Chandra} observatory and to 
constrain the electron energy distribution by measuring
the X--ray fluxes produced by this effect.
We also discuss the evidence for  
IC scattering of nuclear photons in powerful 
radio galaxies.
\end{abstract}

\section{Introduction}

A full understanding of the energetics and of the energy 
distribution
of the relativistic particles in the radio jets and lobes
of radio galaxies and quasars is of basic importance to 
a complete description of the physics and time evolution 
of these sources. 
Synchrotron emission properties result from a complicated 
convolution of magnetic 
fields and relativistic electron distributions,
so that they cannot be unambiguously derived by 
radio observations alone (see Rudnick, this meeting).
In the lobes of radio galaxies and quasars 
this degeneracy could be broken by measurements of the X-rays
produced by the inverse Compton (IC) scattering of
CMB photons (e.g. Harris \& Grindlay 1979),
and/or nuclear photons (Brunetti, Setti \& Comastri 1997).

While the X--rays produced by the IC scattering of the
CMB photons typically involve the less energetic part of
the radio emitting electrons (Lorentz factor $\gamma \sim 1000$),
those from the IC scattering of the far IR -- optic nuclear
photons are mainly powered 
by $\gamma=100-300$ electrons which
are not detected, since their synchrotron emission falls in the  
hundred kHz to MHz frequency range.
By combining measurements of the 0.1--10 keV extended 
fluxes it is then
possible to obtain broad band information on the energy 
distribution of the relativistic electrons 
($\gamma \sim 50-3000$)
in the lobes of radio galaxies and quasars.

The knowledge of the low energy part of the 
electron spectrum is particularly important.
Indeed, while at higher energies
radiative losses are dominant, 
at lower energies the electron spectrum is 
particularly sensitive to the initial injection conditions
and to the acceleration mechanisms
(Eilek \& Huges 1990).
Furthermore, the detection of X--ray fluxes produced by
IC scattering of nuclear photons, implying the lack of a 
prominent low energy cut off ($\gamma < 300$) in the electron 
energy distribution, may significantly increase the energy budget
in the radio lobes compared with present estimates based on 
standard equipartition formulae (Brunetti et al. 1997).

\section{Anisotropic IC scattering and relativistic 
electron spectrum}

\begin{figure}
\plottwo{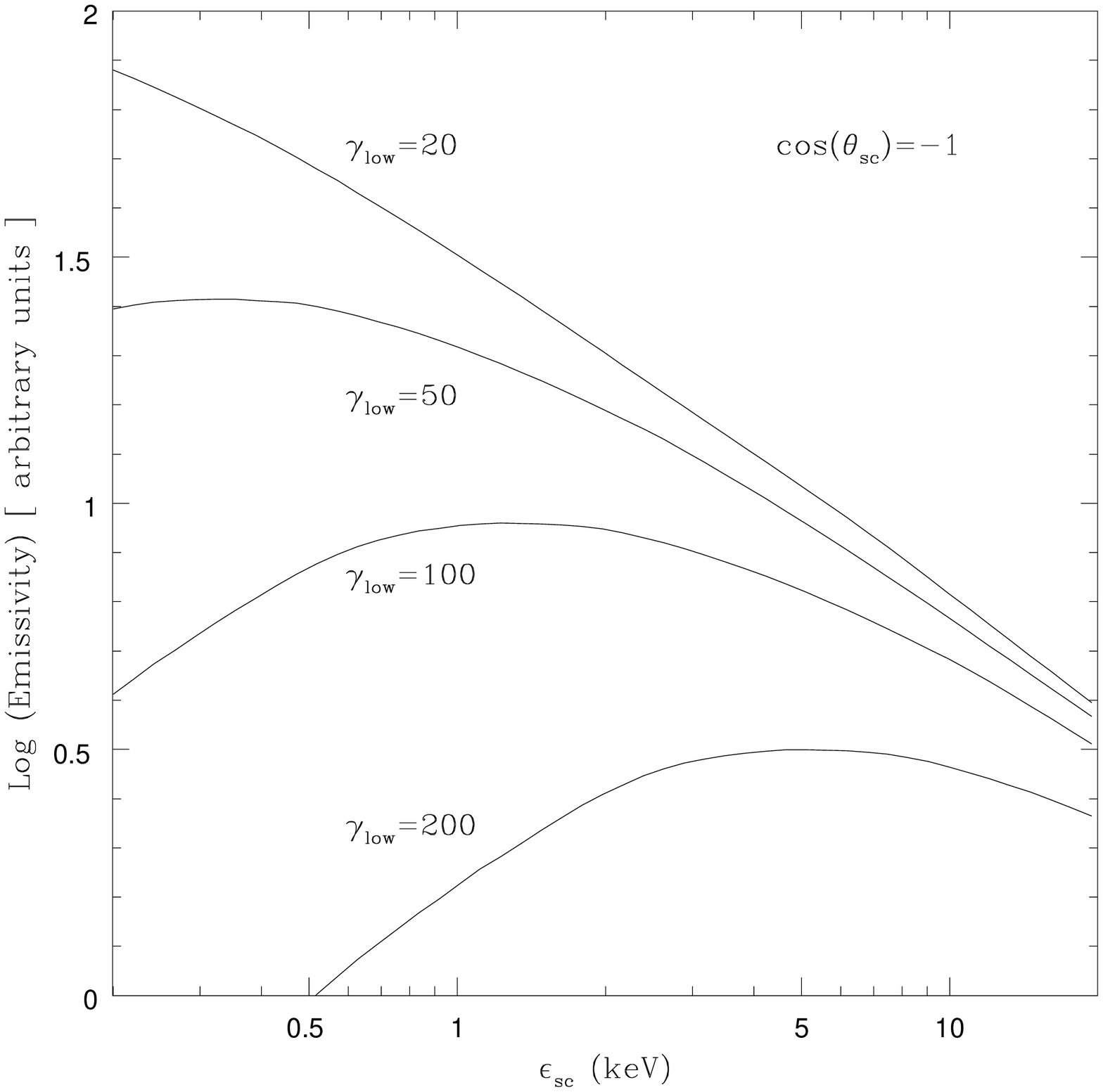}{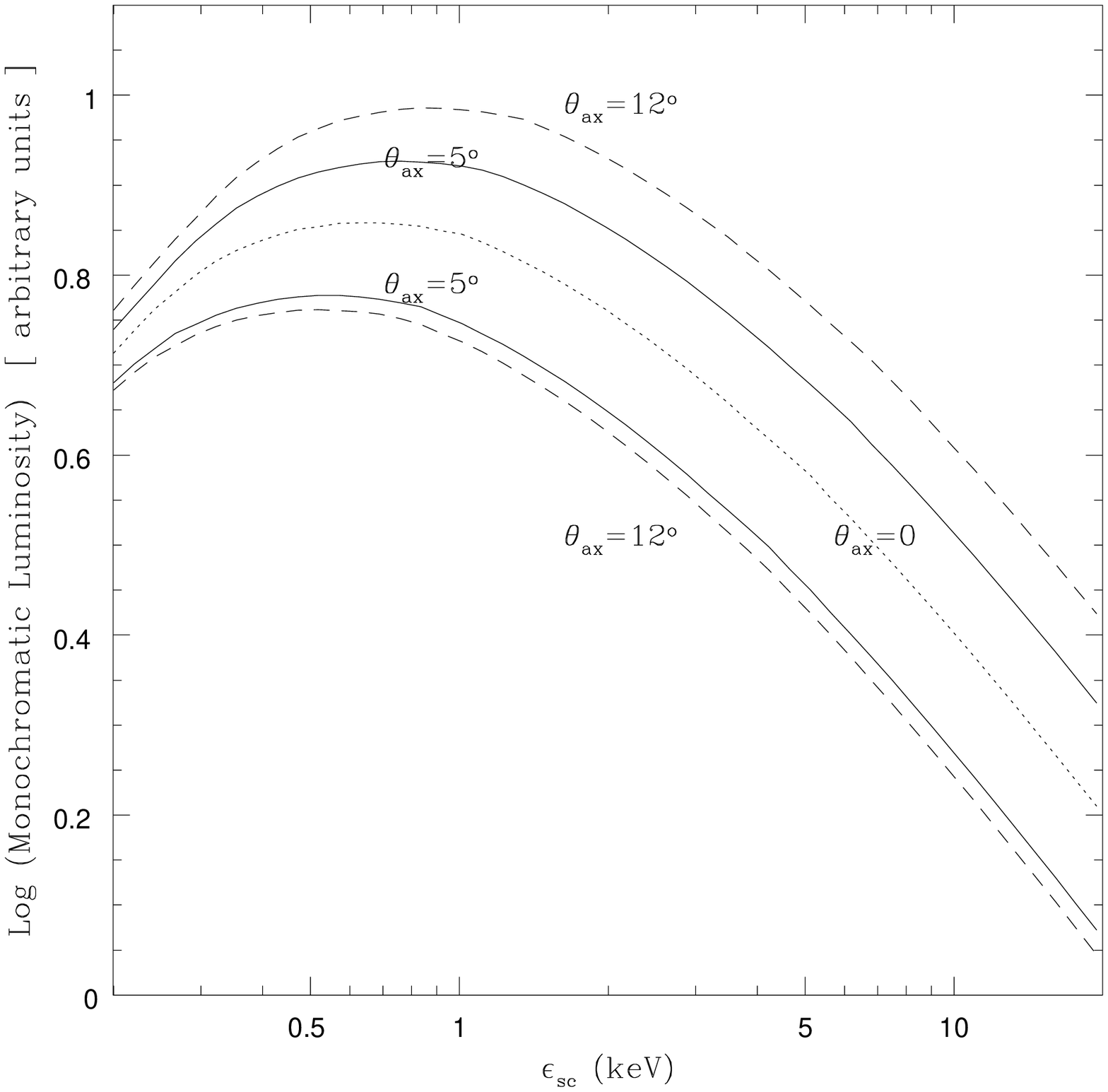}
\caption{a) Spectra from anisotropic IC scattering as a function
of the low energy cut--off
$\gamma_{low}$ in the case of back scattered photons.
In the calculation we assume Sanders 
et al.(1989) SED for the spectrum
of the seed photons and a power law electron energy distribution
of slope 2.5.
b) Spectra from anisotropic IC scattering of the far (upper) and
near lobes as a function of the inclination angle 
($\theta_{ax}$) on the sky plane 
computed in the case of
an ellipsoidal radio galaxy with ratio between major and minor
axis = 4. An opening angle of the quasar radiation cone 
$=40^o$ and $\gamma_{low}=100$ are assumed, the other 
parameters being the same as in panel a).}
\end{figure}

In the framework of the unification between powerful radio galaxies
and radio loud quasars, the IC scattering of far IR to optic/UV
nuclear photons by the relativistic electrons in the radio lobes
could give detectable X--ray fluxes
(Brunetti et al. 1997). 
Due to the geometrical
configuration, the IC scattering is anisotropic so that 
if a symmetric double lobed radio galaxy 
is inclined with respect to the plane of the 
sky, the far lobe should appear more luminous than 
the near one.
The slope of the X--ray spectrum produced by this scattering 
are reasonably close to the radio synchrotron one, 
but, if a low energy flattening of the electron
energy distribution (or a cut--off) is present 
at energies $\gamma < 100$ the spectrum 
should flatten in the low energy part of the
X--ray band (Fig.1a) and, furthermore, since the photons 
from the far lobe are back scattered, 
the X--ray spectrum of the far lobe should appear 
slightly harder that that of the near one 
(Fig.1b).

For a given energy distribution of the scattering electrons, 
the anisotropic IC spectrum can be obtained by making
use of Brunetti (2000) equations.
A positive detection of X--rays generated by this
process can be used to
constrain the electron energy distribution and 
the possible presence of a low energy cut--off ($\gamma_{low}$).
In addition to the brightness asymmetry caused by the
orientation of the radio axis,  
the X--rays from IC scattering of nuclear photons are
expected to be more concentrated in the innermost parts
of the radio lobes due to the dilution of the nuclear
photon flux with distance.
Moreover, if the lobes of a radio galaxy are significantly 
different in shape, the radio lobe closer to the nuclear 
source is expected to be the most luminous.
All these properties of the X--ray brightness distribution 
allow to recognise the X--rays from 
IC scattering of nuclear photons.

\section{Observational evidence}

\subsection{The case of 3C 219}

Possible evidence for this emission has been found in the case 
of the powerful radio galaxy 3C 219 (z=0.1744) by a
relatively deep ROSAT HRI observation corroborated 
by a combinated ROSAT PSPC and ASCA spectral analysis
(Brunetti et al. 1999). 
In this case the 
observed X--ray flux 
can be matched by assuming that the 
magnetic field 
in the inner radio lobes
is a factor of $\sim 3$ lower than the
equipartition value (calculated with $\gamma_{low}=50$), 
similarly to that found in the case of 
Cen B (Tashiro et al.1998).
3C 219 will be observed by {\it Chandra}
and we eagerly expect confirmation
of the IC scenario.

\subsection{The case of 3C 295}

\begin{figure}
\plottwo{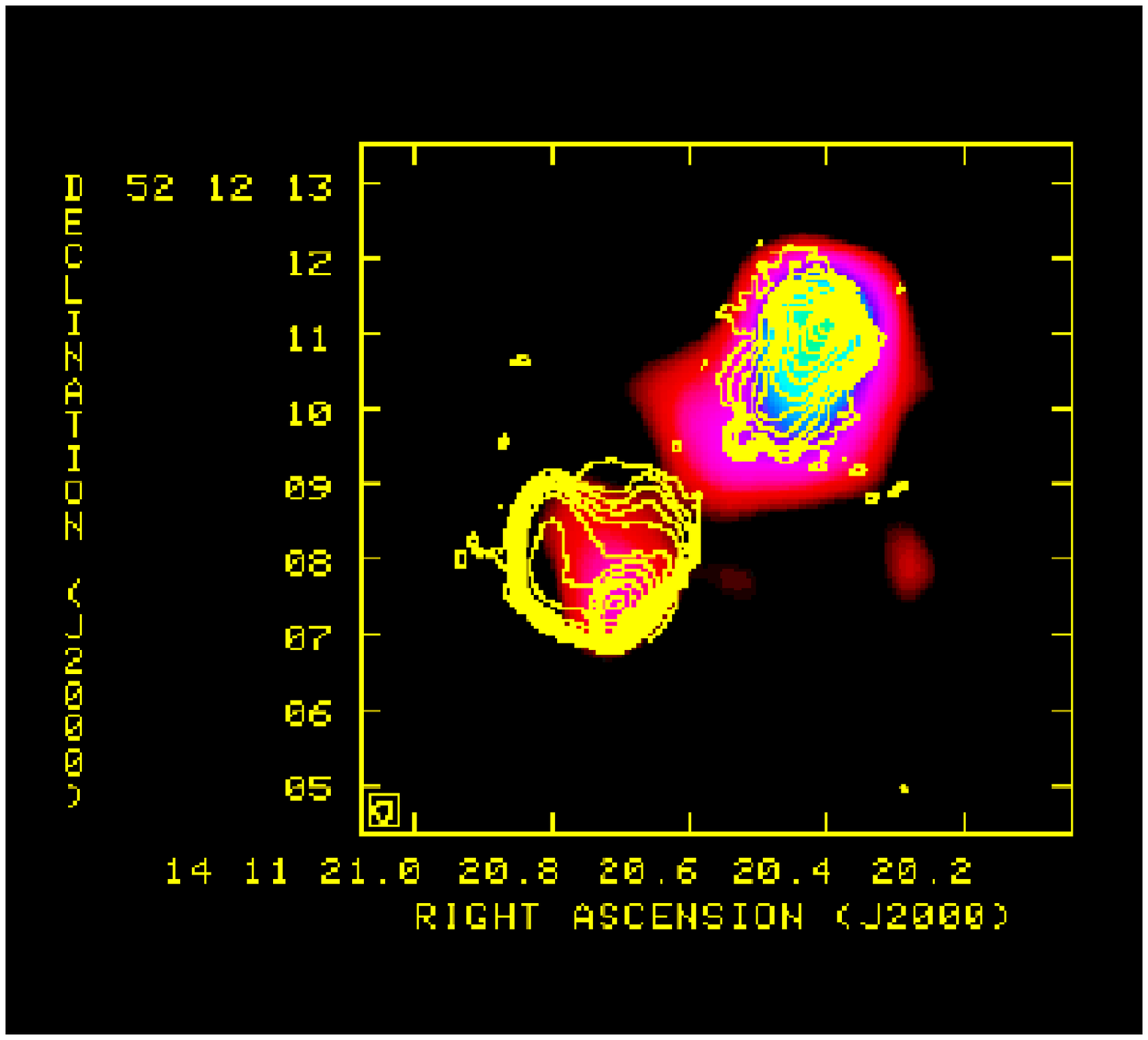}{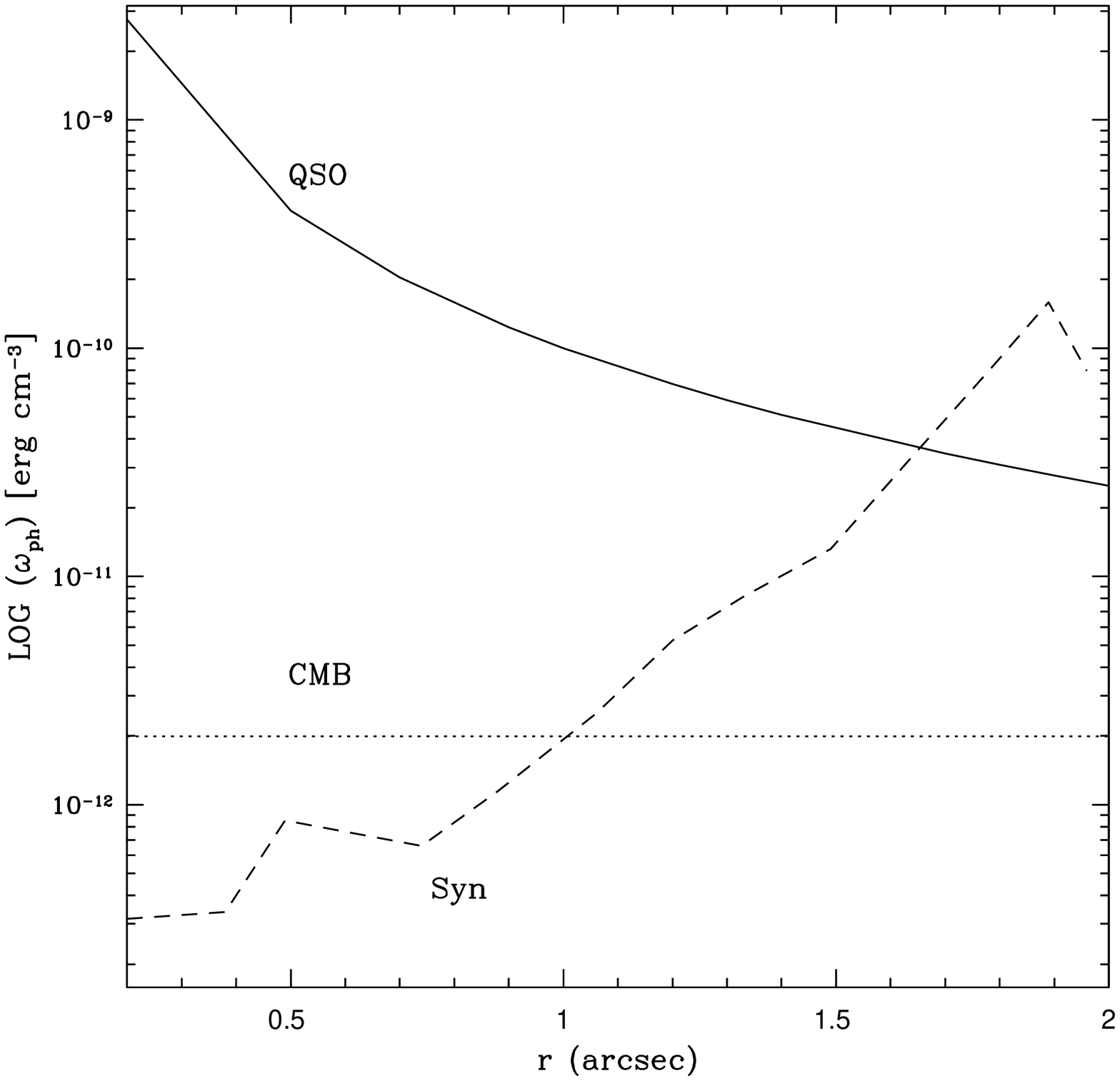}
\caption{a) The X--rays (color) after the cluster subtraction 
are shown superimposed
on the radio image (contours; kindly provided 
by Taylor \& Perley). 
b) The energy density of the
nuclear photons along the radio axis of the
northern lobe of 3C 295 is compared
with those of the synchrotron 
and CMB photons ($L_{qso}=10^{46}$erg s$^{-1}$ is
assumed).}
\end{figure}

We report here on a study 
(Brunetti et al., to be submitted) of 
the powerful and compact FR II radio source 3C 295, 
which extends for $\sim 5$ arcsec on the plane of the sky, 
and is identified with a giant elliptical (cD) galaxy at the 
center of a rich cluster, at redshift z=0.461.
This source was observed by {\it Chandra}
(calibration time) for an elapsed time of $\sim$20 ks;
the 0.5--7 keV image is dominated by the two hot spot regions
and by a powerful nuclear source (Harris et al.2000)
making the study of extended X--rays associated 
with the radio lobes very difficult. 
We have performed a 0.1--10 keV spectral analysis of the nucleus
finding that it is well fitted by a powerful X--ray
source ($\sim 10^{45}$erg s$^{-1}$) absorbed below
$\sim 2$ keV ($N_H \sim 10^{23}$cm$^{-2}$). 
Thus the 0.1-2 keV image is nucleus free allowing to
study the extended 
X--ray emission after subtraction of the 
diffuse cluster component. The result is shown in 
Fig.2a: extended and asymmetric X--ray
emission is clearly detected.
In Fig.2b we report the energy densities due to synchrotron
emission, CMB photons and hidden quasar photons by assuming
a far IR to optic luminosity of $10^{46}$erg s$^{-1}$.
It can be noticed that in the region where extended
X--ray emission is clearly detected the energy 
density due to
the nuclear photons is more than one order of magnitude 
larger than that due to other processes.
By assuming for simplicity that the radio volume is 
axisymmetric, 
an inclination angle with respect to the sky plane of $\sim 5^o$ 
(with the northern being the far lobe) is sufficient to reproduce
the X--ray brightness asymmetry when the different morphologies  
of the two radio lobes are taken into account; such
an asymmetry cannot be explained with other X--ray mechanisms
proposed so far. Furthermore, a faint radio jet 
has been recently discovered in the southern lobe by a deep
MERLIN observation (Leahy, private comm) possibly confirming
that the northern lobe is the farthermost.
Spectral analysis in the 0.1-2 keV band of the extended emission
with a power law model limits the photon index in the range
1.23--1.65 (90\% confidence level) requiring $\gamma_{low} <100$
if a spectral energy injection index $\delta =2.3$ 
(as constrained by radio spectral fits, see Brunetti, 
this meeting) is assumed for the 
electron energy distribution.
By comparing the synchrotron flux from the lobes with the
extended X--ray flux we obtain a magnetic field
$B \sim 40-90 \mu$G (close to the equipartition value, but 
lower) if a reasonable far IR to optical 
luminosity of the hidden quasar of $1-2 \cdot 10^{46}$erg
s$^{-1}$ is assumed.

\acknowledgments

We would like to thank A.Comastri and L.Feretti 
who are actively involved in 
the IC scattering research, and M.Cappi and D.E. Harris who are 
coinvestigators for 3C 295. 
It is also a great pleasure to acknowledge 
Robert Laing and Katherine Blundell
for organizing such an interesting meeting. 
This work was partly supported by the Italian Ministry for
University and Research (MURST) under grant Cofin98-02-32, 
and by the Italian Space Agency (ASI).

\end{document}